\documentclass[aps,superscriptaddress]{revtex4-1}
\usepackage{graphicx}
\usepackage{dcolumn}
\usepackage{bm}
\usepackage{epstopdf}
\begin{document}
%
\title{A stochastic model for non-relativistic particle acceleration }

\author{G. Pallocchia}
\email[]{giuseppe.pallocchia@iaps.inaf.it}
\affiliation{ INAF - Istituto di Astrofisica e Planetologia Spaziali, Via 
del Fosso del Cavaliere 100, 00133 Roma}
\author{M. Laurenza}
\affiliation{ INAF - Istituto di Astrofisica e Planetologia Spaziali, Via 
del Fosso del Cavaliere 100, 00133 Roma}
\author{G. Consolini}
\affiliation{ INAF - Istituto di Astrofisica e Planetologia Spaziali, Via 
del Fosso del Cavaliere 100, 00133 Roma}
\begin{abstract}
A stochastic model is proposed for the acceleration of non-relativistic 
particles yielding to energy spectra with a shape of a 
Weibull\textquoteright s function. Such particle distribution is found 
as the stationary solution of a diffusion-loss equation in the 
framework of a second order Fermi\textquoteright s mechanism producing 
anomalous diffusion for particle velocity.  The present model is supported 
by in 
situ observations of energetic particle enhancements at interplanetary 
shocks, as here illustrated by means of an event seen by STEREO B 
instruments in the heliosphere. Results indicate that the second order 
Fermi\textquoteright s mechanism provides a viable explanation for the 
acceleration of energetic particles at collisioness shock waves.

\end{abstract}


\maketitle



One of the most intriguing and unsolved problems of Astrophysics is 
the particle acceleration to high energies in space plasmas. Fermi's 
acceleration mechanism \citep{Fer49} is a theoretical tool  
extensively  used in astrophysical contexts and also in other 
research fields like Plasma Physics \citep{Micha99} and in the theory of 
dynamical systems \citep{Zasla65,Lich91}. 
The first-order acceleration, a variant of original Fermi\textquoteright s 
mechanism, constitutes the basics for the diffusive shock acceleration 
(DSA) \citep[e.g.,][]{Krym77,Bland78} wherein a particle, repeatedly 
scattered across the shock front, gains energy through head-on collisions 
against the converging downstream and upstream plasma irregularities. 
The DSA naturally produces a power law energy spectrum which is accepted 
to explain the observed cosmic-ray spectrum up to about $10^{15} eV$ 
\citep{Bland87}. 
Hence, the DSA approach has received the most attention to interpret 
particle acceleration at shock waves, although it does not fully 
address several aspects of phenomenon. For instance, the expected 
relationship between the power-law spectral index and the shock compression 
ratio at interplanetary shocks is loose when checked through 
observations \citep{van84}. 
Moreover, observations of solar energetic particle (SEP) events have 
shown that the predicted power law is valid on a limited energy 
interval \citep[e.g.,][]{Mew2005} below a characteristic energy 
where the spectrum has a rollover. An exponential decay was only
heuristically introduced to take into account this feature 
\citep{Ell85}, where the rollover energy is supposed to depend on 
several parameters related to the interplanetary shock \citep{Lee2012}.
On the other hand, stochastic acceleration (SA), also called 
second-order acceleration and based on the original Fermi\textquoteright s 
mechanism, is characterized by an average energy gain due to the 
particle interaction with randomly moving magnetized clouds or 
turbulent fluctuations. The SA has been proposed to play a dominant role 
in many other astrophysical environments where particles can be accelerated 
in a bounded space region such as Radio galaxies \citep{Eil79}, solar 
flares \citep{Petro2004}, the interstellar medium  \citep{Seo94}, supernova 
remnants \citep{Scott75}. 
Few theoretical works suggested tha SA could be important at shock waves
as well \citep{Schli93,Ostr93,Afanasiev2014}, altough this has not been 
tested against observations.
Recent observational studies \citep{Lau2013,Lau2015} have shown 
that SEP spectra, as well as spectra of particles 
accelerated at transient and Corotating Interaction Regions (CIRs) 
shocks, can be succesfully fitted by means of a Weibull\textquoteright s 
function \cite{Weib51}. 
Here we propose a theoretical derivation of such a 
Weibull\textquoteright s spectrum through a leaky-box model 
based on a second-order Fermi\textquoteright s mechanism wherein the 
broadening of energy distribution is slower than mean energy gain.
The good agreement with observations and the overall physical consistency 
of the model (both illustrated in an event of acceleration at 
interplanetary shock), provide evidence that SA can be effective at 
collisionless shock. Hence, the present paper offers a 
scenario alternative to that depicted by DSA which is generally
invoked to account for particle acceleration in 
the shock-related physical contexts. \\
Let us start our model derivation from the classical 
Fermi\textquoteright s scheme in which particles 
are stochastically accelerated in a spatial region by interactions 
with randomly moving magnetic irregularities or turbulent 
fluctuations. 
Moreover, let us assume that scattering is effective in making 
the particles distribution isotropic.
In our model the spatial region is homogenous and, consequently, the 
spatial diffusion is not considered.
The number of particles per unit volume and per unit solid angle
having kinetic energies in the range $E$ to $E+\Delta E$, is then 
expressed as $(4\pi)^{-1}N(E,t)\Delta E$, that is only as a function of the 
time and energy. All of the particles are injected in the acceleration 
process with the same energy $E_{in}$ (henceforth we refer all 
energies to $E_{in}$ for notation convenience, thus $E_{in}= 0 $) 
at constant rate of $q_{in}$ particles per unit volume and time. Particle 
leakage from the acceleration region is taken into account through a 
characteristic time of confinement $\tau$ indipendent from the energy.
Hence the appropriate diffusion-loss equation, expressing the conservation
of the number of particles in energy space, reads 
\citep[e.g.][]{GinSy64,Mill90,Lon94}:
\begin{equation}
\frac{\partial N}{\partial t} =  \frac{\partial (b(E)N)}{\partial E} + \frac{1}{2}\frac{\partial^2{(d(E)N)}}{\partial E^2}- \frac{N}{\tau} + q_{in}\delta (E)
\label{dfe}
\end{equation}
where $b(E)=-\frac{d\langle E \rangle}{dt}$,
$d(E)=\frac{d\langle (\Delta E)^2 \rangle}{dt}$
and $\langle \bullet \rangle$ stands for average over a particles 
ensemble. \\
The four terms on right-hand side account respectively for: 
1) the mean "drift" of the particles in energy space ($b(E)$ represents 
the average acceleration rate); 2) the "broadening" of the particle 
energy distribution (terms 1) and 2) are connected with the stochastic 
nature of the acceleration process); 3) particles leakage from the 
acceleration region; 4) supply from sources of monoenergetic beam of 
fresh particles with energy $E_{in}$. \\
In the framework of second-order Fermi\textquoteright s mechanism it is 
known that anomalous (i.e. nonstandard Brownian) diffusion for 
particles velocity can arise \citep{Bou04,Per07}: 
$\langle |\bm{v} (t)-\bm{v}_{0}|^2 \rangle \sim t^{2\nu}$ 
with $\nu\neq 1/2$ ($\bm{v}_{0}$ is the initial velocity).
For instance, \citet{Bou04} developed two-dimensional minimal 
stochastic model in which particles absorb kinetic energy 
(accelerate) through collisions against magnetic irregularities 
modeled as localized moving scattering centers.
They found for both particle velocity and position an 
anomalous superdiffusive behaviour.
Hence, we assume there exists a non-relativistic implementation of 
Fermi\textquoteright s stochastic mechanism in which the particles undergo 
an anomalous diffusion for velocity yielding to:
\begin{equation}
\langle E(t)^n \rangle \sim (t/\tau)^{n\nu(n)}    
\end{equation}
where $n\nu(n)$ is a concave function of $n$ (i.e. its 
slope continually decreases).
The nonlinearity of $n\nu(n)$ indicates that the probability 
distribution function (PDF) of particle velocity at different times 
is not self-similar, namely a PDF of the form  
$P(|\bm{v}|,t) = t^{-\nu}F(|\bm{v}|/t^{\nu})$ cannot describe the 
anomalous diffusion at all time scales by means of the same value 
of $\nu$. Actually, numerical studies on the motion of tracer 
particles in sandpile \citep{Car99} and in plasma turbulence \citep{Car01} 
show that system finite size effects can determine a breakdown of PDF 
self-similarity characterized by a nearly piecewise linear $n\nu(n)$ 
function with a smaller slope for high $n$ than for low $n$. 
Therefore, we justify the assumption of 
concavity for $n\nu(n)$ as a way to account for finite size effects 
on velocity diffusion in the model (e.g. the finite value of the 
probability per unit time $\tau^{-1}$ for a particle to exit 
from the acceleration process). \\
The relative weight of the second to the first  
term on RHS of Eq.(\ref{dfe}) can be easily estimated, through 
dimensional considerations, by the ratio: 
\begin{equation}
R(\langle E \rangle )=\frac{d(\langle E \rangle )}{b(\langle E \rangle )\langle E \rangle }\sim\frac{\langle E^2 \rangle}{\langle E \rangle ^2}
\end{equation}
where we consider $\langle (\Delta E)^2 \rangle \sim \langle E^2 \rangle$.  
Hence, using Eq.(2)
in Eq.(3) and dropping the bracket notation 
(hereafter no longer necessary), we obtain the scaling law:
\begin{equation}
R(\lambda E) = \lambda^{-2\alpha}R(E)
\end{equation}
where $\alpha = \left [1-\nu(2)/\nu(1) \right ]$ and $\lambda > 0$ is 
a scale factor. \\
Since $\alpha > 0$ due to the concavity of 
$n\nu(n)$, Eq.(4) implies that 
$R(E) \ll 1$ if $E=\lambda E_{\ast} \gg E_{\ast}$, 
being $E_{\ast}$ approximately defined through $R(E_{\ast}) \simeq 1$.
Therefore, in energy regime $E \gg E_{\ast}$, the second term on 
RHS of Eq.(1) can be neglected and the steady state spectrum 
($ \partial N / \partial t\equiv 0$)  
obtained by solving Eq.(1) reduced to more simple form:     
\begin{equation}
N= -\frac{E_{\tau}^\beta}{\beta}\frac{\partial (E^{1-\beta}N)}{\partial E}
\end{equation}
where $E_{\tau}=\langle E(\tau) \rangle $ and $\beta \equiv 1/\nu(1) > 0$. 
A straightforward integration yields:
\begin{equation}
N(E)=A(E/E_{\tau})^{(\beta-1)}\mathrm{e}^{-(E/E_{\tau})^\beta}
\end{equation}
where $A$ is an integration constant. Therefore, the accelerated 
particles are distributed according to the Weibull\textquoteright s 
statistics. \\ 
Transients and corotating shocks are systems where particles are 
assumed to be locally accelerated, as energetic particle enhancements 
\citep[e.g.,][and references therein]{Arm70,Gosl81,Lario2003} are 
usually associated with their passage.
\begin{figure}
	\includegraphics[angle=0,scale=0.9]{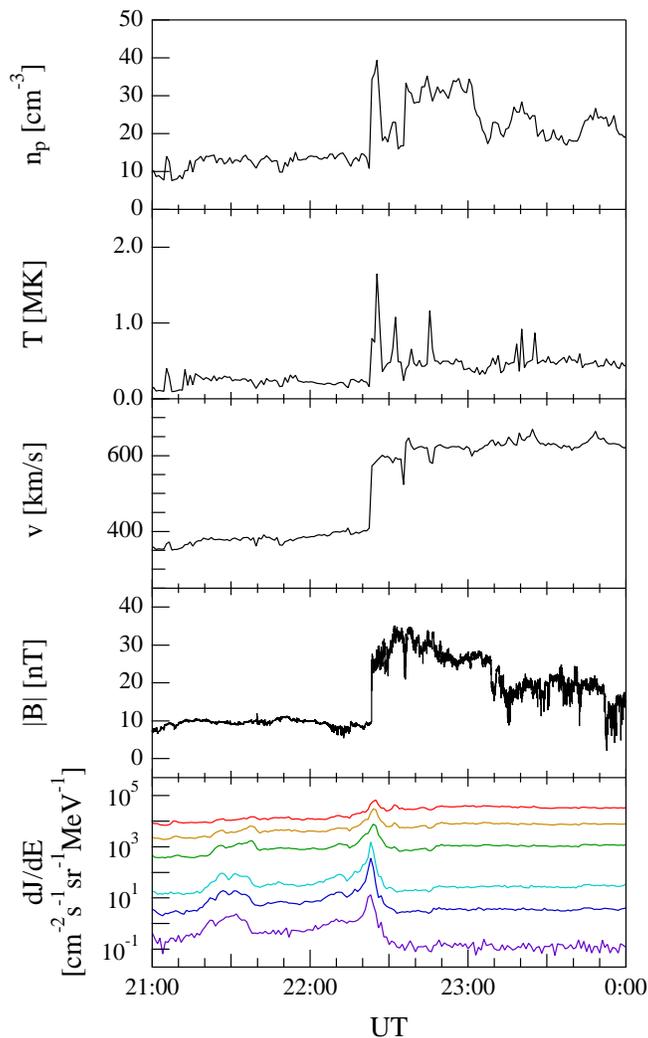}%
	\caption{\label{FiguraFlux} Time history of solar wind plasma 
parameters and energetic particle fluxes as recorded by STEREO B s/c
between 21:00 UT and 24:00 UT on October $3^{rd}$ 2011. From top to 
bottom: the proton density $n_p$ and temperature $T$, bulk speed $v$,  
magnetic field magnitude and the proton differential fluxes for a 
selected number of energy channels ($E \sim 0.53, 1.05, 2.10, 4.74, 6.93$ 
and $10.95$ MeV from top to bottom).}
\end{figure}
Hence, we illustrate the consistency of our model in case the acceleration 
region is a collisionless shock wave. 
We remark that the two fundamental assumptions of the model are 
consistent with physical conditions at interplanetary shock 
where turbulent fluctuations are observed upstream and/or 
dowmstream of the shock front \citep[e.g.,][]{Ken82,Zank06}. 
In fact, from the theoretical point of view, turbulence can provide 
efficient particle scattering (thus supporting the first basic 
assumption of the model)
\citep[e.g.,][]{Tver68, Bland87,Petr2012, Byko2014} 
to account for the isotropy of the observed energetic particle 
distribution function [30]. In addition, it can be responsible 
for momentum diffusion (second basic assumption of the model)
\citep{Tver68, Bou04} so that the energy of the turbulent field 
is transferred to particles through a stochastic Fermi\textquoteright s
mechanism \citep[e.g.,][]{Tver68,Fedo2012,Petr2012}. \\     
On 3 October 2011 at 22:23 UT, STEREO B spacecraft (located at 1.08 
AU, -98.09$^\circ$ and 1.08$^\circ$ heliographic longitude and 
latitude, respectively) observed a quasi-perpendicular fast shock moving 
radially outward from Sun with a speed $v_{sh}\simeq 700~km/s$. 
At the same time, a particle enhancement was recorded by 
the onboard instruments SEPT, LET and HET in the energy range 
$0.1 - 100~MeV$. Figure 1 reports a quicklook of 
the main plasma and particle parameters along with magnetic field 
intensity measurements. 
Data used to study this event are 1 minute averaged proton fluxes 
measured by the three instruments. 
This event occurs on a quiet background and the intensities 
start to rise sharply at the shock passage. 
Figure 1 shows that the proton peak is found at 22:23 UT, when the 
shock can be identified by the abrupt changes in the solar wind 
parameters. \\ 
An average differential flux was calculated on the time interval 
22:14 - 22:31 UT around the shock arrival and a best-fit was performed 
by means of a function derived from Eq.(5) taking into account the 
conversion from the particle spectrum to the differential flux 
($dJ/dE=C \times N(E) \times E^{1/2}$). 
The obtained values for the best-fit parameters are: 
$C\sim [2.0\pm0.5]*10^5 cm^{-2}s^{-1}sr^{-1}MeV^{-1}$, $\beta=[0.50\pm0.07]$ 
and $E_\tau=[95\pm5] KeV$. 
As shown in Figure 2, there is an excellent agreement between our model
and the experimental data over the wide energy range 
$0.3\div 30~MeV$ spanning around two orders of magnitude. 
As expected from the model, the agreement is worst at lower energies.
In turbulent plasmas the theoretical escape time from the acceleration
region due to the spatial transport is $\tau(E)\sim E^{-\gamma}$ 
\citep[e.g.,][]{Fedo2012}. If we assumed such $\tau(E)$ in the 
model, the resulting (softer) spectrum would differ from that 
in Eq.(6) just for the replacement
$(E/E_{\tau})^\beta \rightarrow \beta/(\beta+\gamma)(E/E_{\tau})^{\beta+\gamma}$
in the exponential factor. 
However, we verified in the present case that such a correction to 
Weibull\textquoteright s spectrum ($\gamma=0$) results to be negligible 
for energies lower than several tens of $MeV$. Therefore, in spite of
extreme simplicity, our assumption of constant escape time proves to 
be reasonable by virtue of the good agreement between the present 
leaky-box model and observations. \\
 \begin{figure}
 	\includegraphics{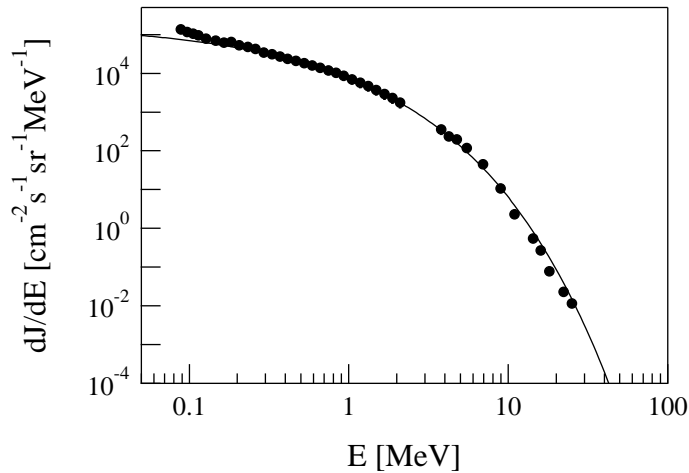}%
 	\caption{\label{FiguraESP} Differential flux averaged over 
the time interval 22:14 - 22:31 UT around the shock arrival on 
October $3^{rd}$ 2011. Black  curve is the best-fit Weibull\textquoteright s 
function. Data errors are within the marker size.}
 \end{figure}
In our model $\tau$, $\beta$ and $E_{\tau}$ are free parameters which 
can assume, in principle, any value independently from each other. 
We show that the obtained estimates are congruent with a physical 
picture of the event. The value of $\beta=0.5$ (viz $\nu(1) = 2$) 
implies superdiffusion for velocity. In general terms, a high degree 
of persistence of the anomalous diffusion is expected for an efficient 
particle acceleration. 
Moreover, as already mentioned, the same superdiffusive behaviour 
spontaneously arises in a minimal model of second order 
Fermi\textquoteright s acceleration proposed by \citet{Bou04}. 
Thus, the above $\beta$ value proves to be fairly meaningful 
from a physical point of view. \\
In case of efficient energization, the mean energy $E_\tau$ gained 
in a characteristic time $\tau$ has to be much higher than the 
typical injection energy. As matter of fact, $E_\tau=95 ~KeV$ 
considerably exceeds both typical bulk flow 
$E_{bulk}=1/2m_{p}{V_{sw}}^2\sim 5~KeV$ and thermal 
$E_{th}=K_{B}T_{p}\sim0.15~KeV$ energies of the upstream solar 
wind protons (see Fig.1). 
Hence, it is consistent with the reasonable hypothesis that the energetic
particle population is accelerated directly out of the ambient solar wind. \\
The confinement time $\tau$  cannot be directly obtained through the 
best-fit procedure. Nevertheless, observations can provide upper and 
lower limits for its value.
In fact, taking into account $\beta = 0.5$ and $E_\tau=95~KeV$, it is seen 
from Eq.(2) that a particle energy of $\sim 30~MeV$ (viz the highest energy
in Fig.2) is reached after a time $T_{high}\simeq 18\tau$. 
Obviously, $T_{high}$ can equal, at most, the shock travelling time from 
the Sun to the spacecraft position $R_{s/c} = 1.08 AU$, that is 
$T_{stt}= R_{s/c}/v_{sh}\simeq 2.7~days$. Hence, the upper limit is
$\tau_{up} \simeq 3.6~hr$. On the contrary, in case of nearly local 
acceleration, $T_{high}$ must be of order of the time width of particle 
enhancement that, in present case, is around $10 \div 20~min$. The 
lower limit is, therefore, $\tau_{low}\simeq 1~ min$.   
When calculated from Eq.(2) with the above values of $\beta$, $E_{\tau}$ 
and $\tau$, the acceleration time scales of our superdiffusive model 
result to be comparable with DSA ones or even shorter. 
For instance, \citet{Zha13} estimate that DSA accelerates a proton 
to an energy of $\sim 10 ~MeV$ in a time of $\sim 12 ~hr$ at $1 ~AU$ 
(see their Fig.1). In our model, the same energy is reached after 
a time $\sim 10\tau$ which may range from $\sim 10 ~min$ to $\sim 36 ~hr$ 
depending on the actual $\tau$ value. 
It is conceivable that a second order Fermi\textquoteright s acceleration
may be more efficient than DSA.
For instance, \citet{Ostr94a} has showed 
that, under the hypothesis of negligible damping of very low frequency 
Alfv\'{e}n waves, statistical acceleration by high-amplitude MHD 
turbulence can transfer the energy of a weak parallel shock to the 
particles more efficiently than a first order process.  
Moreover, \citet{Schli93} proposed that due to efficient momentum 
diffusion of particles in the downstream region of the shock, the 
acceleration can be dominated by the second-order acceleration mechanism. \\
In summary, we have introduced a simple stochastic model to obtain a 
Weibull\textquoteright s spectrum for accelerated energetic particles. 
The fundamental assumption was that acceleration is given by an 
anomalous diffusion in momentum space characterized by a 
broadening of the energy distribution slower than average energy gain.
Afterwards, through the analysis of an event registred in the 
interplanetary space, we showed that the model can account for 
the observations at collisionless shock over a wide energy range 
and that its acceleration time scales are competitive with those of 
the diffusive shock acceleration (DSA). \\
In conclusion, the present study is particularly important since it 
provides evidence that a second order Fermi\textquoteright s process 
may efficiently accelerate particles at shock waves, viz, in a 
physical environment where instead DSA is usually thought to 
play the dominant role.
Moreover, we point out that the parameters of the 
Weibull\textquoteright s spectrum  acquire a clear physical 
meaning within our model and, hence, their experimental estimates represent 
a helpful tool in interpreting the observations of energetic particles 
connected with several solar and interplanetary phenomena such as SEP, 
CIR and transient collisionless shocks.    
Nevertheless, further theoretical and observational efforts are needed to 
better understand the details of the microphysics of the magnetic 
field turbulence around the collisionless shock front and how it can 
affect the trapping and acceleration of energetic 
particles \citep[e.g.,][]{Verkh05}.   
%
%
%
%
%
%
\begin{acknowledgments}
The authors would like to thank all of STEREO teams for making data 
available.
This research has been supported by the European Community's Seventh 
Framework Programme ([FP7/2007-2013]) under Grant agreement 
no. 313038/STORM and ASI/INAF contract no. I/022/10/0.
\end{acknowledgments}
%
\providecommand{\noopsort}[1]{}\providecommand{\singleletter}[1]{#1}%

\end{document}